\documentclass[
reprint,
 amsmath,amssymb,
 aps,
 pra,
]{revtex4-2}

\usepackage{graphicx}
\usepackage{dcolumn}
\usepackage{bm}

\begin{document}
\preprint{APS/123-QED}

\title{Ion Stochastic Heating by Low-frequency Alfv\'en Wave Spectrum}

\author{Jingyu Peng}
\author{Jiansen He}%
 \email{jshept@pku.edu.cn}
\affiliation{%
 School of Earth and Space Sciences, Peking University, Beijing 100871, Beijing, China
}%




\date{\today}

\begin{abstract}
Finite-amplitude low-frequency Alfv\'en waves are commonly found in plasma environments, such as space plasmas, and play a crucial role in ion heating. The nonlinear interaction between oblique Alfv\'en wave spectra and ions has been studied. As the number of wave modes increases, ions are more likely to exhibit chaotic motion and experience stochastic heating.
The stochastic heating threshold in the parameter space can be characterized by a single parameter, the effective relative curvature radius $P_{{eff.}}$. The results show excellent agreement with the chaotic regions identified through test particle simulations. The anisotropic characteristics of stochastic heating are explained using a uniform solid angle distribution model. The stochastic heating rate $Q=\dot{T}$ is calculated, and its relationship with wave conditions is expressed as $Q/(\Omega_i m_i v_A^2) = H(\alpha) \tilde{v}^3 \tilde{B}_w^2 \tilde{\omega}_1$, where $\alpha$ is propagating angle, $\Omega_i$ is the gyrofrequency, $m_i$ is the ion mass, $v_A$ is the Alfv\'en speed, $\tilde{v}$ is the dimensionless speed, $\tilde{B}_w$ is the dimensionless wave amplitude, and $\tilde{\omega}_1$ is the lowest dimensionless wave frequency.
\end{abstract}

\maketitle


\section{\label{sec:intro}introduction}

The heating of ions in the corona and solar wind is a critical topic in  heliospheric physics. Observations show that ion heating in these regions typically displays two key characteristics: preferential heating of heavy ions and  heating predominantly in the perpendicular direction~\cite{Kohl1998,Li1998,Antonucci2000,Telloni2007,Dolei2015,Marsch1982,Marsch2004,Hellinger2006,Tracy2016,Kasper2017,Peng2024,Stansby2019}. Alfv\'en waves (AWs) resonating with ions near the ion gyrofrequency $\Omega_i$ are considered a primary mechanism for ion heating, as cyclotron resonance heating naturally accounts for both perpendicular heating and preferential heating of heavy ions~\cite{Gary2001,Gary2006,Tu2001,Cranmer2001,Hollweg2002}. However, the role of  cyclotron resonance in heating  coronal and solar wind ions remains uncertain, since there is no direct evidence that such high-frequency waves possess the energy required to heat ions. Observations indicate that most Alfv\'en wave (AW) energy in the corona is concentrated at low frequencies~\cite{Jess2009,Chashei2000,DePontieu2007,Tomczyk2007}, and similarly, solar wind AW turbulence energy is  predominantly found at large scales~\cite{Smith1995,Tu1995,Bale2005,Bruno2013}. Moreover, due to the nature of perpendicular turbulent cascades, the transfer of wave energy to higher frequencies is highly inefficient~\cite{Cranmer2003,Howes2008,Sahraoui2009,Chandran2010}.

Some studies have found that, even without satisfying the cyclotron resonance condition, low-frequency AWs still exhibit certain wave-particle interaction mechanisms that enable ion heating, such as pickup, phase randomization, and stochastic heating. First, the pickup and heating of ions by low-frequency AWs have been investigated in Refs.~\cite{Dong2013,Dong2014,Wang2006,Wang2009,Lu2022,Wu2023,Lu2007}. This heating process consists of two components. The first involves the heating of  newborn ions, which are picked up by the waves with different initial phases, resulting in varying magnetic moments. The second component arises from motion that is parasitic on the waves, specifically the $\bm{E} \times \bm{B}$ drift caused by the  wave electric field $\bm{E_w}$ and the background magnetic field $\bm{B_0}$. When the wave dissipates adiabatically, this component of  heating disappears and is therefore referred to as pseudo-heating~\cite{Wang2009,Dong2013}. The heated ions exhibit motion on two time scales: the ion gyrofrequency and the wave frequency~\cite{Lu2009,Sun2014,Wu2023}. Ultimately, the change in ion temperature is proportional to the wave energy,
\begin{equation}
T^{p.u.}=T_{0}+m_{i}v_{A}^{2}\frac{B_w^2}{B_0^2},
\end{equation}
where $T^{p.u.}$ is the final temperature resulting from the pickup mechanism, $T_{0}$ is the initial temperature, $m_{i}$ is the mass of ion species $i$, $B_0$ is the background magnetic field, $B_w$ is the wave amplitude, $v_{A}=\frac{B_0}{\sqrt{\mu_0\rho_{m}}}$ is the Alfv\'en speed, $\mu_0$ is the vacuum magnetic permeability, and $\rho_{m}$ is the plasma mass density. The pickup process is completed within one gyro-period~\cite{Dong2014}.

Second, if ions have different initial velocities, their parallel thermal motion causes phase randomization~\cite{Sun2014,Nariyuki2010,Lu2007,Lu2009,Wu2023,Li2007}, which leads to further heating with a characteristic time $t=\pi/kv_{th}$, where $k$ is the wave number and $v_{th}$ is the initial thermal speed~\cite{Lu2007,Li2007}.

Third, when the wave amplitude exceeds a certain threshold,  particle motion becomes chaotic, allowing stochastic heating by low-frequency waves~\cite{Chen2001}. Ion stochastic heating caused by low-frequency monochromatic AWs with various polarization relations has been investigated~\cite{Chen2001,Lv2007,Sun2014,Kolesnychenko2005}, and the Poincar\'e surface of section (PSOS) is used to distinguish between regular and chaotic trajectories in the state space. The interaction between multiple low-frequency AW modes and ions is studied~\cite{Kolesnychenko2005,Lu2009}. The stochastic heating threshold decreases as the number of  wave modes increases. However, PSOS becomes inapplicable for identifying stochastic heating  in the presence of multiple wave modes. Therefore, the ion velocity power spectrum is used to assist in this determination, as demonstrated in Ref.~\cite{Lu2009}. The heating rate of stochastic heating is analyzed in Ref.~\cite{Sun2014}, which reports that stochastic heating exhibits a timescale of 10–20 minutes, with the heating rate increasing linearly with  wave energy density, frequency, and propagation angle, while it decreases with increasing plasma $\beta$.

A recent paper proposes quantifying chaos using the maximum Lyapunov exponent $\lambda_m$~~\cite{Lyapunov1992,Benettin1980,Geist1990,DynamicalSystems.jl-2018,DatserisParlitz2022} and the Chaos Ratio $CR$~\cite{peng2025}. The authors highlight that the physical picture of ion chaotic motion induced by low-frequency AWs is the breakdown of magnetic moment conservation, which results from wave-induced field line curvature (WFLC). They further explain that the onset of chaos can be determined by the effective relative curvature radius
\begin{equation}
P_{eff.}=\frac{R_c||\nabla \mathbf{B}||_F}{\rho_i||\nabla_{\perp}\mathbf{B}||_{F}},
\end{equation}
where $\bm B$ is the magnetic field, $||\cdot||_F$ is the Frobenius norm, $\rho_i=\frac{m_iv_\perp}{q_i|\bm B|}$ is the gyro-radius, $v_\perp$ is the velocity perpendicular to the magnetic field, $q_i$ and $m_i$ are the charge and mass of species $i$, respectively, and $R_c=\left|\bm{B}\right|^2/\left|\bm{B}\cdot\nabla\bm{B}\right|$ is the field line curvature radius. However, only monochromatic waves are considered. In this paper, we extend the method for quantifying chaos to wave spectra and demonstrate that the condition $P_{\text{eff}} < C$ (with $C \approx 25$) remains valid for the breakdown of magnetic moment conservation and the onset of chaos. Moreover, the criteria for chaos are more easily met in the case of wave spectra. We also calculate the stochastic heating rate for quasi-perpendicular AW spectra and offer a qualitative explanation for the anisotropic heating.

In this paper, we start by examining single-particle trajectories and then extend our analysis to the trajectories of many particles. This approach enables us to explore temperature and heating—that is, the average effects across a large number of particles. The paper is organized as follows: Section~\ref{sec:method} introduces the governing equations of the system and the generation of wave spectra. In Section~\ref{subsec:particleMotion}, we analyze single-particle trajectories. Section~\ref{subsec:heating} addresses the trajectories and heating of numerous particles, while Section~\ref{subsec:heatingRate} investigates the stochastic heating rate. Finally, we present our conclusions and discussion in Section~\ref{sec:dis}.

\section{\label{sec:method}method}

The fundamental assumptions and governing equations follow those presented in Ref.~\cite{peng2025}. The number of wave modes is $N$. To reduce the degrees of freedom in the parameters, we assume that all wave modes share the same wave vector direction within the $x-z$ plane, with a propagating angle $\alpha=\angle(\bm{k},\,\bm{B_0})=\arctan(k_x/k_z)$, where $\bm{B_0}=B_0\hat{\bm{e_z}}$ is the constant background magnetic field. Their frequencies in the plasma frame $\omega_k$ are uniformly distributed within the range $[\omega_1, \,\omega_1 + 0.08\Omega_i]$. According to the dispersion relation, $k_z = \omega_k/v_A$. The amplitude $B_k$ of the $k$th wave mode is related to its frequency $\omega_k$
\begin{equation}
B_k^2/B_0^2=(\omega_k/\omega_1)^{-q},\,q=1.667.
\end{equation}
The sum of the squared amplitudes of all wave modes is $B_w^2=\sum_kB_k^2$. The total magnetic field
\begin{equation}
\bm{B} = \bm{B_0}+\bm{B_w},
\end{equation}
where
\begin{eqnarray}
\bm{B_w}&=&\sum_k B_k[-\cos(\alpha)\sin(\psi_k)\hat{\bm{e_{x}}}+\cos(\psi_k)\hat{\bm{e_{y}}}\nonumber\\
&+&\sin(\alpha)\sin(\psi_k)\hat{\bm{e_{z}}}],
\end{eqnarray}
is the wave magnetic field, and $\psi_k=\bm{k}\cdot\bm{x}+\phi_k$ is the phase of the $k$th wave mode.

The analysis in this paper is conducted in the wave frame, and the governing equation of motion for an ion of species $i$ is~\cite{Chen2001}
\begin{subequations}
\label{eq:ODE}
\begin{equation}
\dot{\psi}_k=k_xv_x+k_zv_z,\label{subeq:ODE_a}
\end{equation}
\begin{equation}
\dot{\bm{v}}=\Omega_i\bm{v}\times\left({\bm{\hat{z}}}+{\bm{B_w}}/{B_0}\right),\label{subeq:ODE_b}
\end{equation}
\end{subequations}
\begin{equation}
    \dot{\bm{x}}=\bm{v},
\end{equation}
where the gyrofrequency $\Omega_i=q_iB_0/m_i$. Eq.~\ref{eq:ODE} form a complete ordinary differential equation (ODE) system describing the motion of ions in a $(N+3)$-dimensional dimensionless state space $\bm{s}=(\psi_1,\,...,\,\psi_N,\,\tilde{v_x}, \,\tilde{v_y}, \,\tilde{v_z})=(\psi_1,\,...,\,\psi_N,\,v_x/v_A, \,v_y/v_A, \,v_z/v_A)$. This ODE system involves $3$ dimensionless parameters: $\tan\alpha$, $\tilde{B_w}^2=B_w^2/B_0^2$ and $\tilde{\omega_1}=\omega_1/\Omega_i$. Notably, since the particles are influenced solely by the Lorentz force, the dimensionless velocity $\tilde{v}=v/v_A$ remains constant.

\section{\label{sec:res}results}

\subsubsection{\label{subsec:particleMotion}Single Particle Motion}

\begin{figure}[htb]
\includegraphics[scale=0.0691]{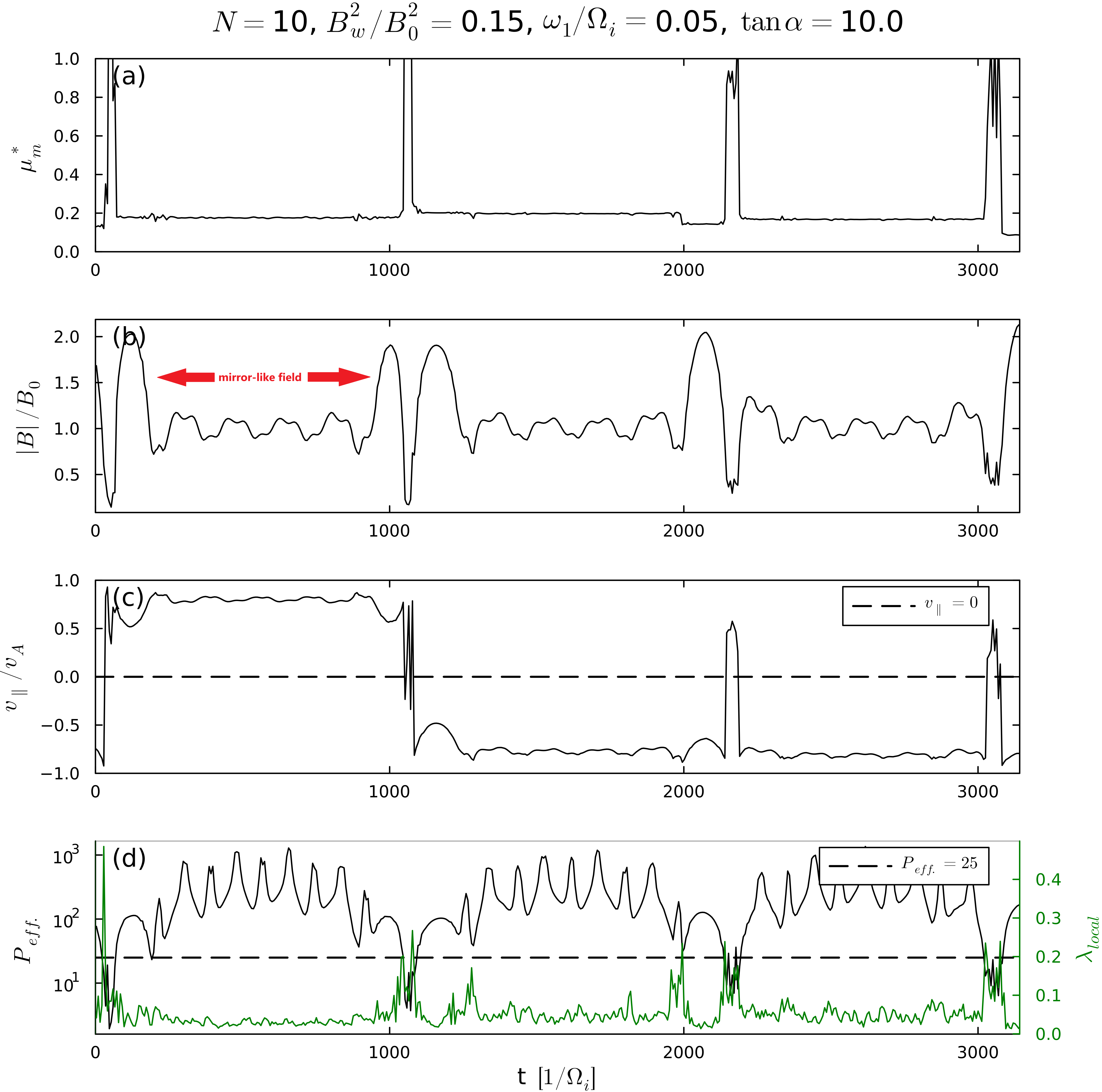}
\caption{\label{fig:singleParticleTimeSeries}Time series of chaotic motion under wave conditions $B_w^2/B_0^2=0.15,\,\omega_1/\Omega_i=0.05,\,\tan\alpha=10$ and $N=10$. The particle's initial state is $(0,\,...,\,0,\,0,\,0,\,-1)$. (a) $\mu_m^*$. (b) Magnitude of the magnetic field $|\bm B|/B_0$, a mirror-like field is marked. (c) Parallel velocity ${v_{\parallel}}=\bm{v}\cdot{\langle\bm{B}\rangle_{\Omega_i}}/{\left|\langle\bm{B}\rangle_{\Omega_i}\right|}$, the dashed line indicates the positions where velocity reverses, i.e., where $v_\parallel=0$. (d) Black line: effective relative curvature radius $P_{eff.}$, with a dashed line indicates $P_{eff.}=25$. Green line: the maximum local Lyapunov exponent $\lambda_{local}$ calculated over one gyro-period.}
\end{figure}
\begin{figure}[htb]
\includegraphics[scale=0.082]{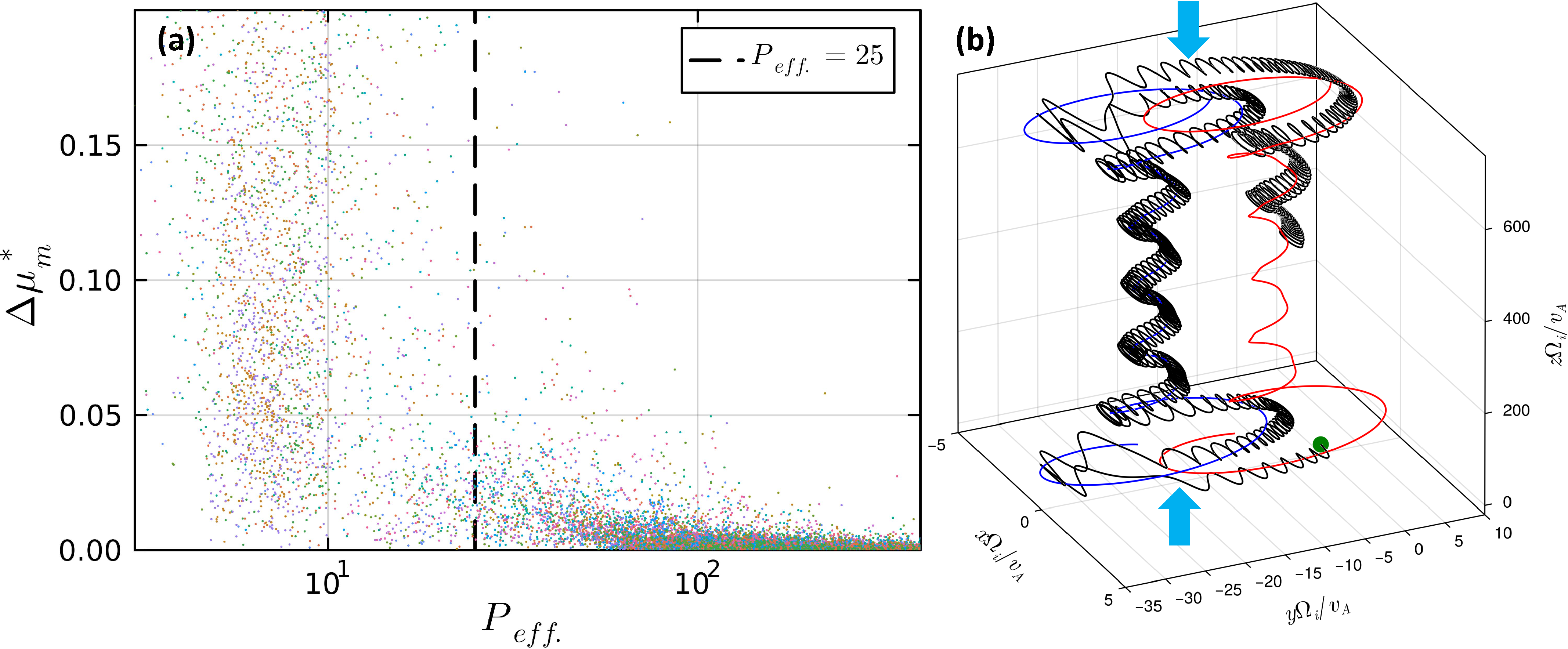}
\caption{\label{fig:muChange}(a) The change in $\mu_m^*$ between neighboring gyro-periods $\Delta\mu_m^*=\frac{1}{2}\left(\left|\mu_m^*-\mu_{m,-1}^*\right|+\left|\mu_m^*-\mu_{m,+1}^*\right|\right)$ at different $P_{eff.}$, where $\mu_{m,-1}^*$ and $\mu_{m,+1}^*$ represent the values of $\mu_m^*$ of the previous and next gyro-periods, respectively. Different colors represent particles with different initial states. A total of 50 particles are considered, each with speed $v=v_A$, initial pitch-angles $\theta_0=\angle(\bm{v_0},\,\bm{B_0})$ uniformly distributed in $[0, \pi]$, initial azimuth angle $\phi_0=\arctan(v_{y0}/v_{x0})=0$, and initial phases $\psi_{k0}$ uniformly distributed in $[0, 2\pi]$. The dashed line indicates $P_{eff.}=25$. Wave conditions are the same as those in Fig.~\ref{fig:singleParticleTimeSeries}. (b) The particle's trajectory (black line) and magnetic field lines (red and blue lines) at the period corresponding to the mirror-like field marked in Fig.~\ref{fig:singleParticleTimeSeries}(b). The particle start from the positions marked by the green dot. The 2 light blue arrows mark the position where $\mu_m^*$ changes.}
\end{figure}

The magnetic moment is the first adiabatic invariant of a charged particle moving in a magnetic field~\cite{Jackson1998}. We calculated the dimensionless magnetic moment $\mu_m^*$ during the particle's motion (see Fig.~\ref{fig:singleParticleTimeSeries}(a)), 
\begin{equation}
    \mu_{m}^*=\frac{\frac{1}{2}\langle|\bm{v_{\perp}|}^2\rangle_{\Omega_i}/v_A^2}{\left|\langle\bm{B}\rangle_{\Omega_i}\right|/B_0}\label{eq:mu_m},
\end{equation}
where the perpendicular velocity $\bm{v_{\perp}}=\bm{v}-v_\parallel\frac{\langle\bm{B}\rangle_{\Omega_i}}{\left|\langle\bm{B}\rangle_{\Omega_i}\right|}$, the parallel velocity ${v_{\parallel}}=\bm{v}\cdot\frac{\langle\bm{B}\rangle_{\Omega_i}}{\left|\langle\bm{B}\rangle_{\Omega_i}\right|}$, and $\langle \cdot {\rangle }_{\Omega_i}=\frac{\int_{t}^{t+T_{\Omega_i}}{\cdot }dt'}{T_{\Omega_i}}$ denotes an average over one gyro-period $T_{\Omega_i}=2\pi/\Omega_i$. When the magnetic field changes slowly (see Fig.~\ref{fig:singleParticleTimeSeries}(b)), the magnetic moment remains constant. However, when the magnetic field changes rapidly, the magnetic moment also varies, causing the particle's motion to display significant discontinuities, as shown in Fig.~\ref{fig:singleParticleTimeSeries}(c). The regions where the magnetic moment changes correspond to areas where $P_{eff.}\lesssim 25$, as shown in Fig.~\ref{fig:singleParticleTimeSeries}(d) and Fig.~\ref{fig:muChange}(a).

When multiple wave modes are present, the geometric structure of the magnetic field can become highly complex. The  magnetic field lines form large spirals intertwined with smaller spirals, as shown by the red and blue lines in Fig.~\ref{fig:muChange}(b). Fig.~\ref{fig:muChange}(b) shows the ion trajectories for $t < 1571/\Omega_i$ in Fig.~\ref{fig:singleParticleTimeSeries}. At locations where the magnetic field line curvature is high, which marked by the 2 light blue arrows in Fig.~\ref{fig:muChange}, the condition for conservation of $\mu_m^*$ is broken. The ion can no longer maintain its gyro-motion around the original magnetic field line and instead deviates from this motion until it is recaptured by a new magnetic field line. This change in magnetic moment displays chaotic behavior, as evidenced by the maximum local Lyapunov exponent $\lambda_{local}$~\cite{Abarbanel1991} in Fig.~\ref{fig:singleParticleTimeSeries}(d) and as discussed in Ref.~\cite{peng2025}. The particle shown in Fig.~\ref{fig:muChange}(b) exhibits a bounce motion because the magnetic field forms a magnetic mirror, with stronger fields on both sides and a weaker field in the center (see Fig.~\ref{fig:singleParticleTimeSeries}(b)). This behavior is indicated by $v_{\parallel}$ crossing zero (see Fig.~\ref{fig:singleParticleTimeSeries}(c)).

\begin{figure*}[htb]
\includegraphics[scale=0.17]{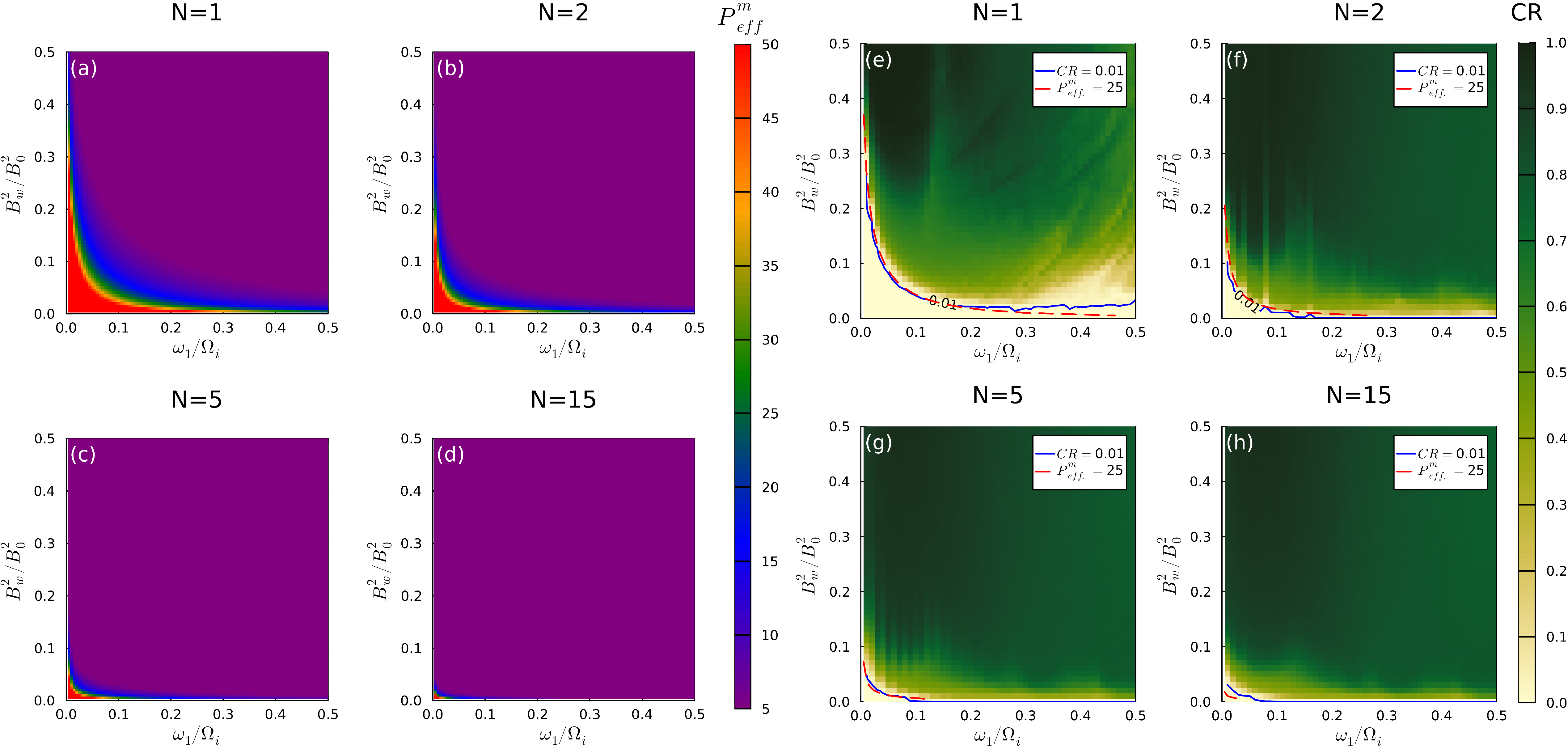}
\caption{\label{fig:Peffmin_CR}$P_{eff.}^m$ and $CR$ in the parameter space $(\omega_1/\Omega_i,\,B_w^2/B_0^2)$, $\tan\alpha=5$. (a)-(d) $P_{eff.}^m$ in the parameter space, with $N=1,\,2,\,5,\,15$. (e)-(h) $CR$ and contour lines of $CR$ and $P_{eff.}^m$ in the parameter space for $N=1,\,2,\,5,\,15$. The $CR$ calculation considered $2500$ particles with $v=v_A$, $\theta_0$ uniformly distributed in $[0, \,\pi]$, $\phi_0 = 0$, and $\psi_{k0}$ randomly distributed in $[0,\, 2\pi]$. The blue lines represent the $CR = 0.01$ contour, while the red lines correspond to the $P_{eff}^m = 25$ contour.}
\end{figure*}

The criterion for determining the change in $\mu_m^*$ in a monochromatic AW is $Peff < C,\, (C \approx 25)$~\cite{peng2025}. This criterion also holds for multiple wave modes, as shown in Fig.~\ref{fig:singleParticleTimeSeries}(d) and Fig.~\ref{fig:muChange}(a). The minimum value of $P_{eff.}$ is
\begin{equation}
    P_{eff.}^m(B_k,\,\mathbf{k})=\frac{\Omega_i}{v B_0}\min_{\psi_k}\left[\frac{|\mathbf{B}|^3}{|\mathbf{B}\cdot\nabla\mathbf{B}|}\frac{||\nabla\mathbf{B}||_F}{||\nabla_{\perp}\mathbf{B}||_F}\right].
\end{equation}
For a monochromatic AW,
\begin{equation}
    \frac{|\mathbf{B}|^3}{|\mathbf{B}\cdot\nabla\mathbf{B}|}\frac{||\nabla\mathbf{B}||_F}{||\nabla_{\perp}\mathbf{B}||_F}=\frac{\left(1+B_w^{*2}+2B_w^{*}\sin\alpha\sin\psi\right)^{3/2}}{k_zB_w^*\sin\alpha},
\end{equation}
reaches its minimum value at a phase of $\psi_m=\frac{3}{2}\pi+2n\pi,\,n\in Z$. For multiple wave modes, expressing $P_{eff.}^m$ analytically is challenging, as it requires identifying the minimum value over the entire phase space $(\psi_1,\,...,\,\psi_N)$. We address this by employing gradient descent.
\begin{figure}[htb]
\includegraphics[scale=0.025]{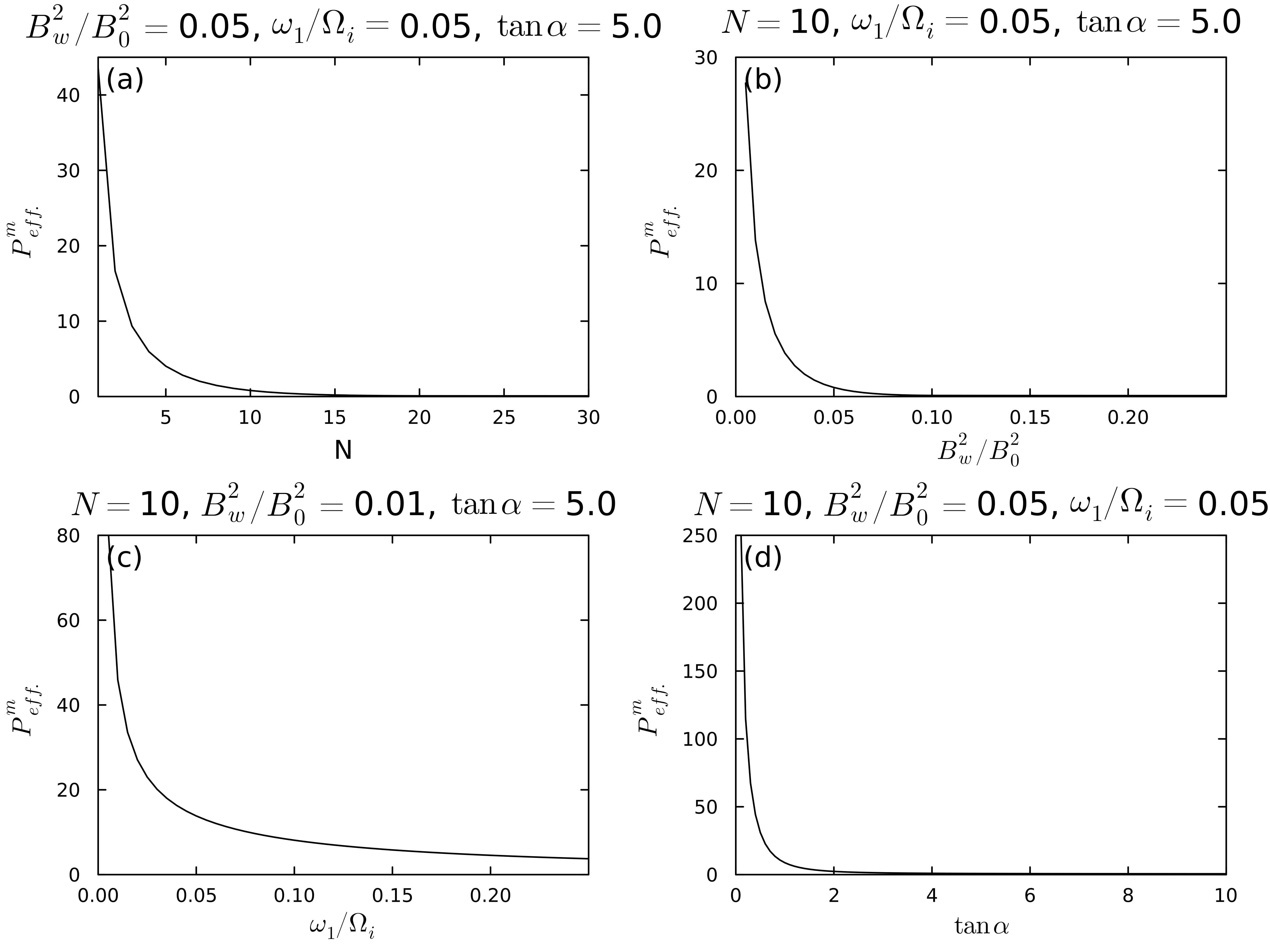}
\caption{\label{fig:Peffmin_params}$P_{eff.}^m$ varies with wave conditions: (a) $N$ (b) $B_w^2/B_0^2$ (c) $\omega_1/\Omega_i$ (d) $\tan\alpha$.}
\end{figure}
As shown in Fig.~\ref{fig:Peffmin_params}, $P_{eff.}^m$ decreases rapidly with increasing $N$, $B_w^2/B_0^2$, $\omega_1$, and $\tan\alpha$, indicating that the system becomes more prone to chaos as these parameters and $N$ increase. In Fig.~\ref{fig:Peffmin_CR}, we plot $P_{eff.}^m$ in the parameter space and also calculate the Chaos Ratio $CR$ (i.e., the ratio of particles exhibiting chaotic motion  among those with different initial states, see Ref.~\cite{peng2025} for further details) using test particle simulations. Chaotic regions are identified where $CR > 0.01$. The chaotic border defined by $P_{eff.}^m = 25$ aligns closely with the results from the test particle simulations. As $N$ increases to 15, global chaos occurs in nearly the entire parameter space $(\omega_1/\Omega_i,\,B_w^2/B_0^2)$. Our findings suggest that the conditions for chaos are  easily met in the continuum spectrum ($N\to\infty$) of space plasma.

\subsubsection{\label{subsec:heating}Heating}

\begin{figure}[htb]
\includegraphics[scale=0.045]{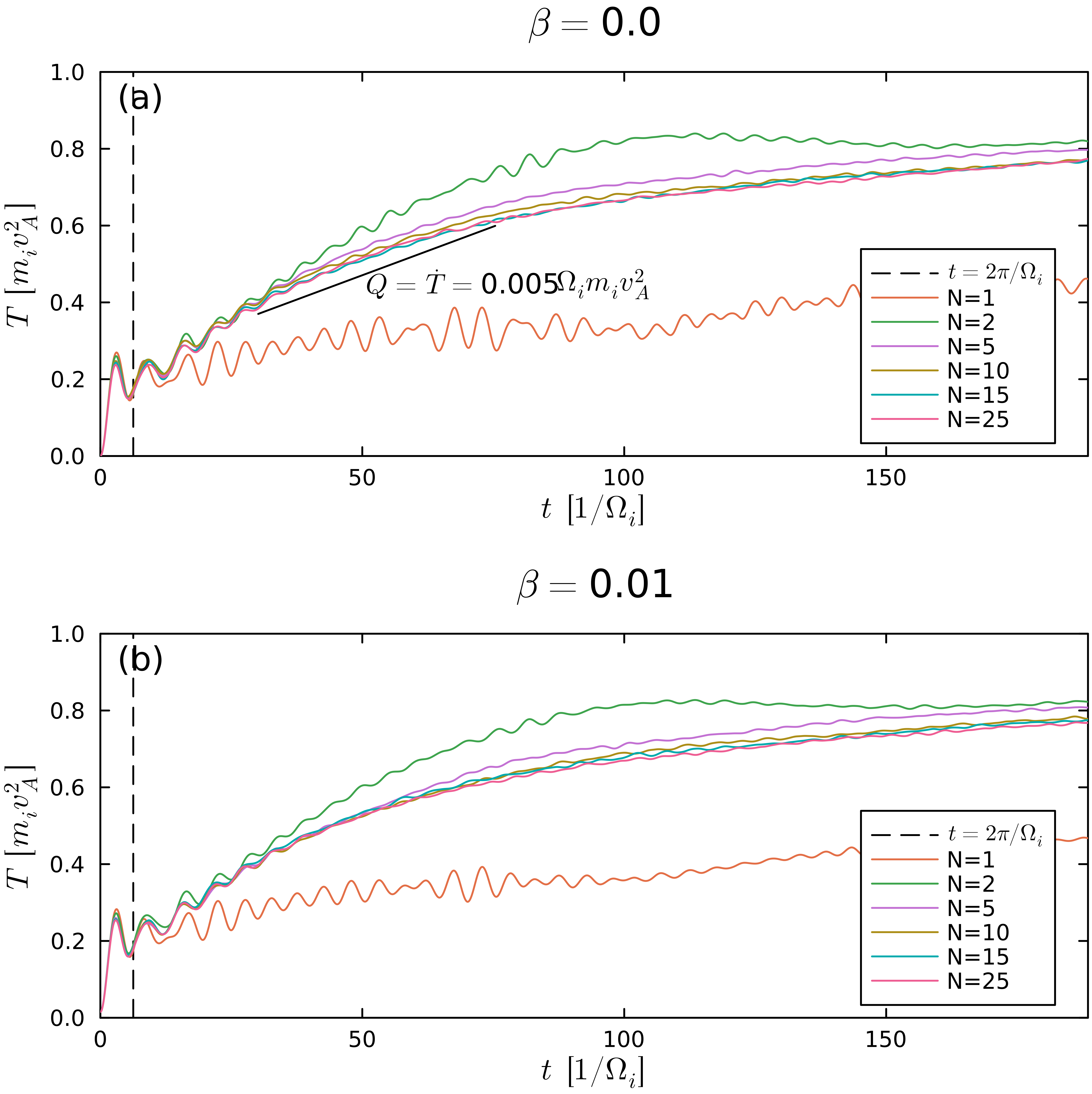}
\caption{\label{fig:temp_time}Ion temperature $T$ varies over time with the parameters $B_w^2/B_0^2=0.15$, $\omega_1/\Omega_i=0.1$, $\tan\alpha=5$. The vertical dashed lines indicate the time $t = 2\pi/\Omega_i$. Solid lines in different colors correspond to different $N$. (a) Cold plasma, $\beta = 0$. The black solid line represents the heating rate $Q=\dot{T}$ for $N=10$, which is calculated from data within the time interval $[2\pi/\Omega_i,\,100/\Omega_i]$. (b) Low-$\beta$ plasma, $\beta = 0.01$.}
\end{figure}

\begin{figure}[htb]
\includegraphics[scale=0.052]{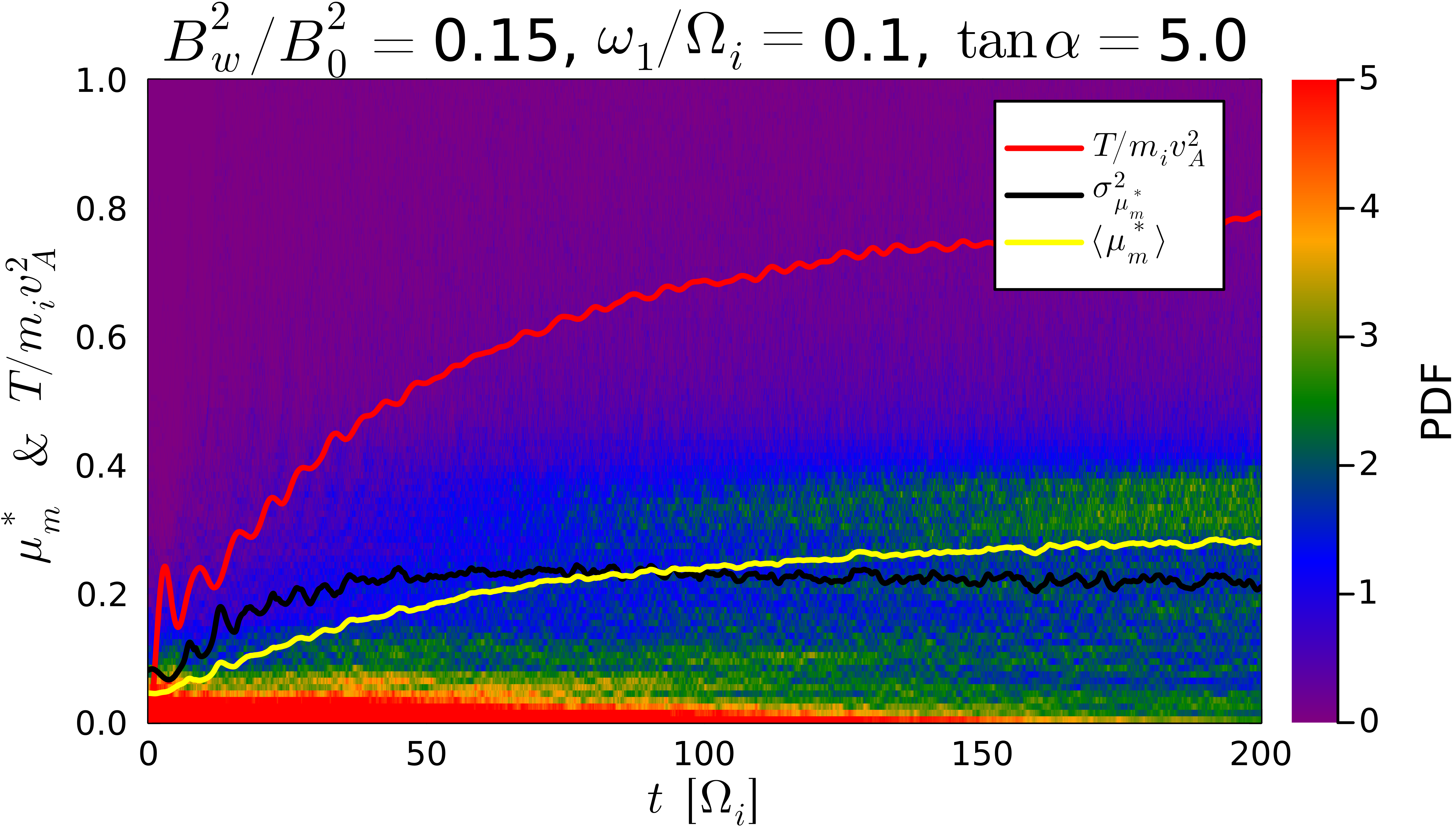}
\caption{\label{fig:mu_time}$T$ and the distribution of $\mu_m^*$ change over time, using the same parameters as in Fig.~\ref{fig:temp_time} with $N=10$. The color indicates the probability density of $\mu_m^*$. Red line: $T$; black line: the variance of $\mu_m^*$; yellow line: the average value of $\mu_m^*$.}
\end{figure}

To investigate the heating of ions by low-frequency AWs, 2000 particles with random initial phases are considered. They possess an initial Maxwellian distribution, with initial thermal speed $v_{th}=\sqrt{\beta}v_A$ and initial bulk velocity $\bm{v_0}=-v_A\hat{\bm{e_z}}$. We then calculate the kinetic temperature, 
\begin{equation}
    T={{m}_{i}}\langle {{\left ({v-\langle v\rangle }\right )}^{2}}\rangle ={{m}_{i}}\left ({\langle {{v}^{2}}\rangle -\langle v{{\rangle }^{2}}}\right ),
\end{equation}
as shown in Fig.~\ref{fig:temp_time}.
When $t < 2\pi/\Omega_i$, the primary heating mechanism is pickup. At $t = 2\pi/\Omega_i$ (indicated by the dashed lines in Fig.~\ref{fig:temp_time}), the system reaches a temperature of approximately $T^{p.u.} \approx m_i v_A^2 \left( \frac{B_w^2}{B_0^2} + \beta \right)$, consistent with theoretical predictions~\cite{Dong2014}.
After $t > 2\pi/\Omega_i$, the temperature increase is primarily due to the stochastic heating mechanism. This heating arises from the diffusion of the magnetic moment distribution, as shown in Fig.~\ref{fig:mu_time}.
The time series of $T$ converges as $N$ increases (see Fig.~\ref{fig:temp_time}). Therefore, in the following analysis, we use $N=10$ to represent the heating characteristics of the continuum spectrum. Test particle simulations indicate that the temperature evolution of low-$\beta$ plasmas (Fig.~\ref{fig:temp_time}(b)) closely matches that of cold plasmas (Fig.~\ref{fig:temp_time}(a)). Consequently, we focus on calculating the heating characteristics for cold plasmas in the subsequent sections. The stochastic heating rate $Q=\dot{T}$ is calculated and shown as the black solid line in Fig.~\ref{fig:temp_time}(a), and the slope is calculated over the interval $[2\pi/\Omega_i,\,t_1]$, where $t_1$ is determined using the method for detecting linear scaling regions described in Ref.~\cite{Datseris2023}.

We investigate the anisotropic characteristics of stochastic heating.
\begin{figure}[htb]
\includegraphics[scale=0.023]{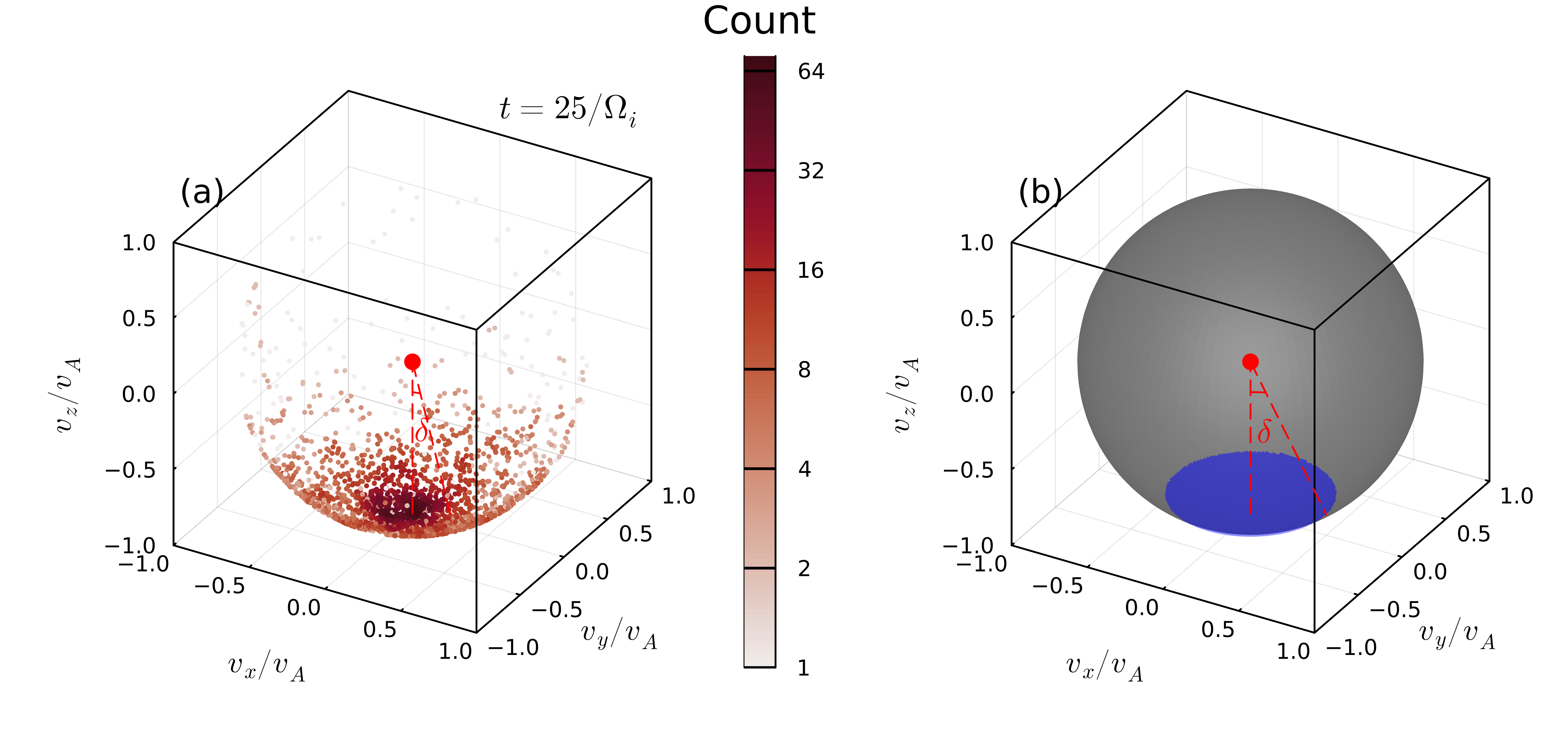}
\caption{\label{fig:spherical_model}(a) Velocity distribution of particles at $t=25/\Omega_i$, where the color indicates the number of points within an $\epsilon$-neighborhood ($\epsilon=\pi/50$) around each point. The wave conditions and the initial states are the same as those in Fig.~\ref{fig:mu_time}. (b) Schematic diagram of the uniform solid angle distribution model, where particles are assumed to be uniformly distributed within the blue region on the gray spherical surface, which has a semi-apex angle of $\delta$. }
\end{figure}
Fig.~\ref{fig:spherical_model}(a) shows the velocity distribution at time $t=25/\Omega_i$. Since the particle speed $v$ remains constant, the velocity distribution always lies on a spherical surface. The stochastic heating process can thus be interpreted as diffusion of the velocity distribution along this spherical surface, i.e., diffusion of the pitch-angle distribution. 
We simplify the stochastic heating process using a uniform solid angle distribution model to qualitatively capture its anisotropic characteristics. Assume that at time $t$, the maximum angle between the velocity $\bm v$ and the $-\hat{\bm{e_z}}$ direction is $\delta=\max\left(\angle(\bm{v},\,-\hat{\bm{e_z}})\right)$, with the velocity  uniformly distributed over solid angles (see the blue region in Fig.~\ref{fig:spherical_model}(b)). Under these assumptions, the temperature $T$, perpendicular temperature $T_\perp$ and parallel temperature $T_\parallel$ can be expressed as follows:
\begin{eqnarray}
    T&=&m_i\left(v^2-\langle v\rangle^2\right)\nonumber\\
    &=&\frac{3-\cos^2\delta-2\cos\delta}{4}m_iv^2,\label{eq:tempModel_T}\\
    T_{\perp}&=&\frac{m_i}{2}\left(\langle v_x^2+v_y^2\rangle-\langle v_x\rangle^2-\langle v_y\rangle^2\right)\nonumber\\
    &=&\frac{-\cos^2\delta-\cos\delta+2}{6}m_iv^2,\\
    T_{\parallel}&=&m_i\left(\langle v_z^2\rangle-\langle v_z\rangle^2\right)\nonumber\\
    &=&\frac{\cos^2\delta-2\cos\delta+1}{12}m_iv^2.
\end{eqnarray}
The temperature change rates can be expressed as
\begin{eqnarray}
    \dot{T}&=&\frac{\sin\delta \dot{\delta}}{2}(1+\cos\delta)m_iv^2,\\
    \dot{T}_\perp&=&\frac{\sin\delta\dot{\delta}}{6}(1+2\cos\delta)m_iv^2,\\
    \dot{T}_\parallel&=&\frac{\sin\delta\dot{\delta}}{6}(1-\cos\delta)m_iv^2\label{eq:hrModel_Tz}.
\end{eqnarray}
\begin{figure}[htb]
\includegraphics[scale=0.034]{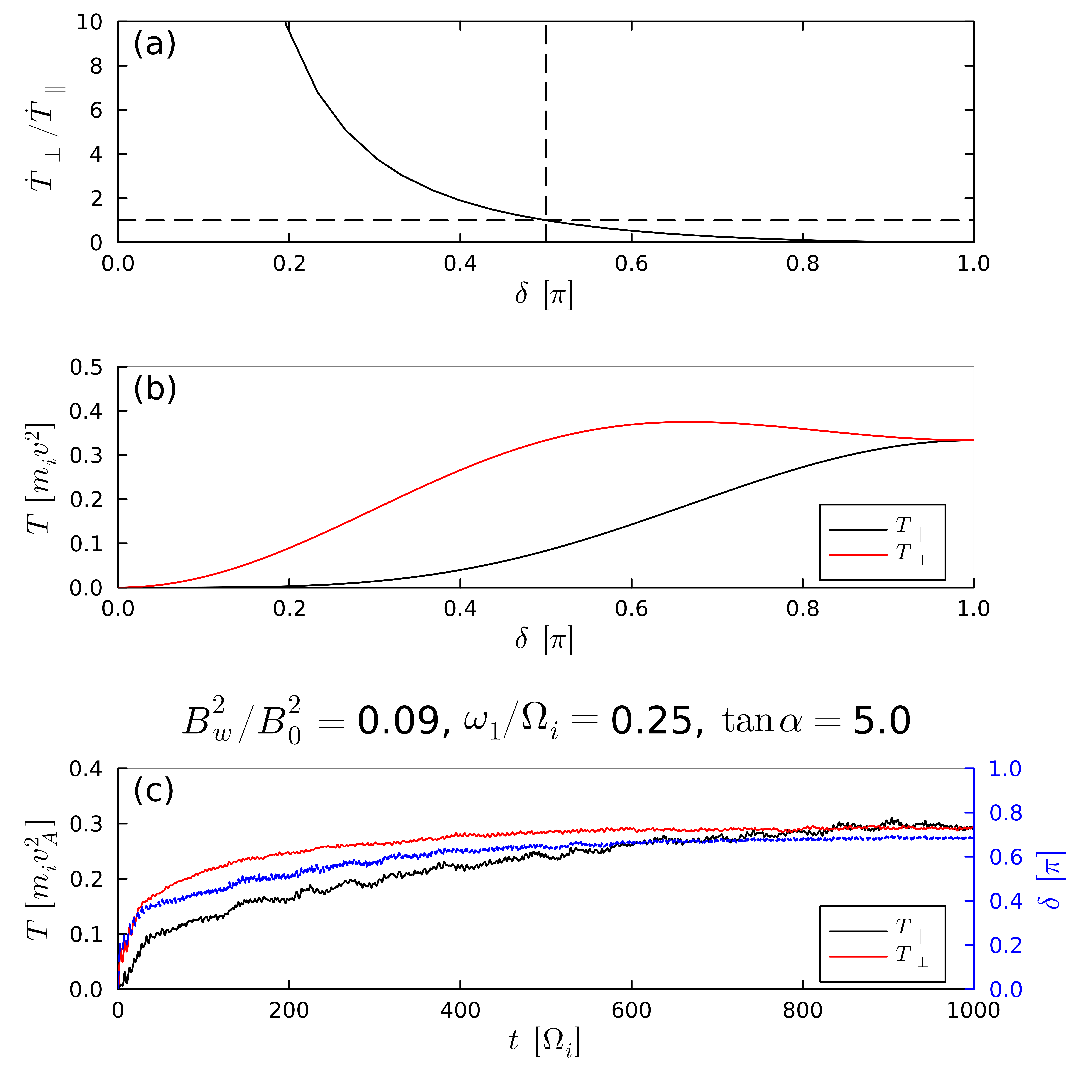}
\caption{\label{fig:temp_model}(a) $\dot{T}_\perp/\dot{T}_\parallel$, (b) $T_\perp$ and $T_\parallel$ as functions of $\delta$ based on the uniform solid angle distribution model. The horizontal dashed line in (a) represents $\dot{T}_\perp/\dot{T}_\parallel=1$, while the vertical dashed line represents $\delta=\pi/2$. (c) Test particle simulation results of $T_\parallel$ (red line) and $T_\perp$ (black line), with $N=10$, $B_w^2/B_0^2=0.09$, $\omega_1/\Omega_i=0.25$, and $\tan\alpha=5$. The blue line indicates the 90th percentile of the angle $\angle(\bm{v},\,-\hat{\bm{e_z}})$.}
\end{figure}
We plot the ratio of perpendicular to parallel heating rates $\dot{T}_\perp/\dot{T}_\parallel$ (Fig.~\ref{fig:temp_model}(a)), as well as the temperatures $T_\perp$ and $T_\parallel$ (Fig.~\ref{fig:temp_model}(b)) as functions of $\delta$, as given by Eq.~\ref{eq:tempModel_T}-\ref{eq:hrModel_Tz}. 
As particles diffuse across the spherical surface, $\delta$ increases from 0 to $\pi$. At $\delta = \pi/2$, with particles uniformly distributed over the hemispherical surface where $v_z\le0$, the stochastic heating is isotropic, i.e., $\dot{T}_\perp/\dot{T}_\parallel=1$, as noted by Ref.~\cite{Sun2014}. When $\delta < \pi/2$,  $\dot{T}_\perp/\dot{T}_\parallel>1$, the stochastic heating preferentially heats the perpendicular direction. Conversely, when $\delta > \pi/2$,  $\dot{T}_\perp/\dot{T}_\parallel<1$, the stochastic heating preferentially heats the parallel direction. 
Fig.~\ref{fig:temp_model}(c) shows the time evolution of $T_\perp$, $T_\parallel$, and $\delta$ based on test particle simulations. Here, $\delta$ is estimated using the 90th percentile of $\angle(\bm{v},\,-\hat{\bm{e_z}})$. Initially, $\delta$ is small, and heating predominantly occurs  in the perpendicular direction. As $\delta$ rapidly increases to $\pi/2$, the heating becomes isotropic. When $t > 200/\Omega_i$, the difference between $T_\perp$ and $T_\parallel$ decreases, indicating preferential heating in the parallel direction. The temperature evolution observed in the test particle simulation in Fig.~\ref{fig:temp_model}(c) qualitatively agrees with the model (i.e., Fig.~\ref{fig:temp_model}(b)), confirming that the anisotropic heating characteristics are dictated by the spherical geometry.

\subsubsection{\label{subsec:heatingRate}Stochastic Heating Rate}

\begin{figure}[htb]
\includegraphics[scale=0.024]{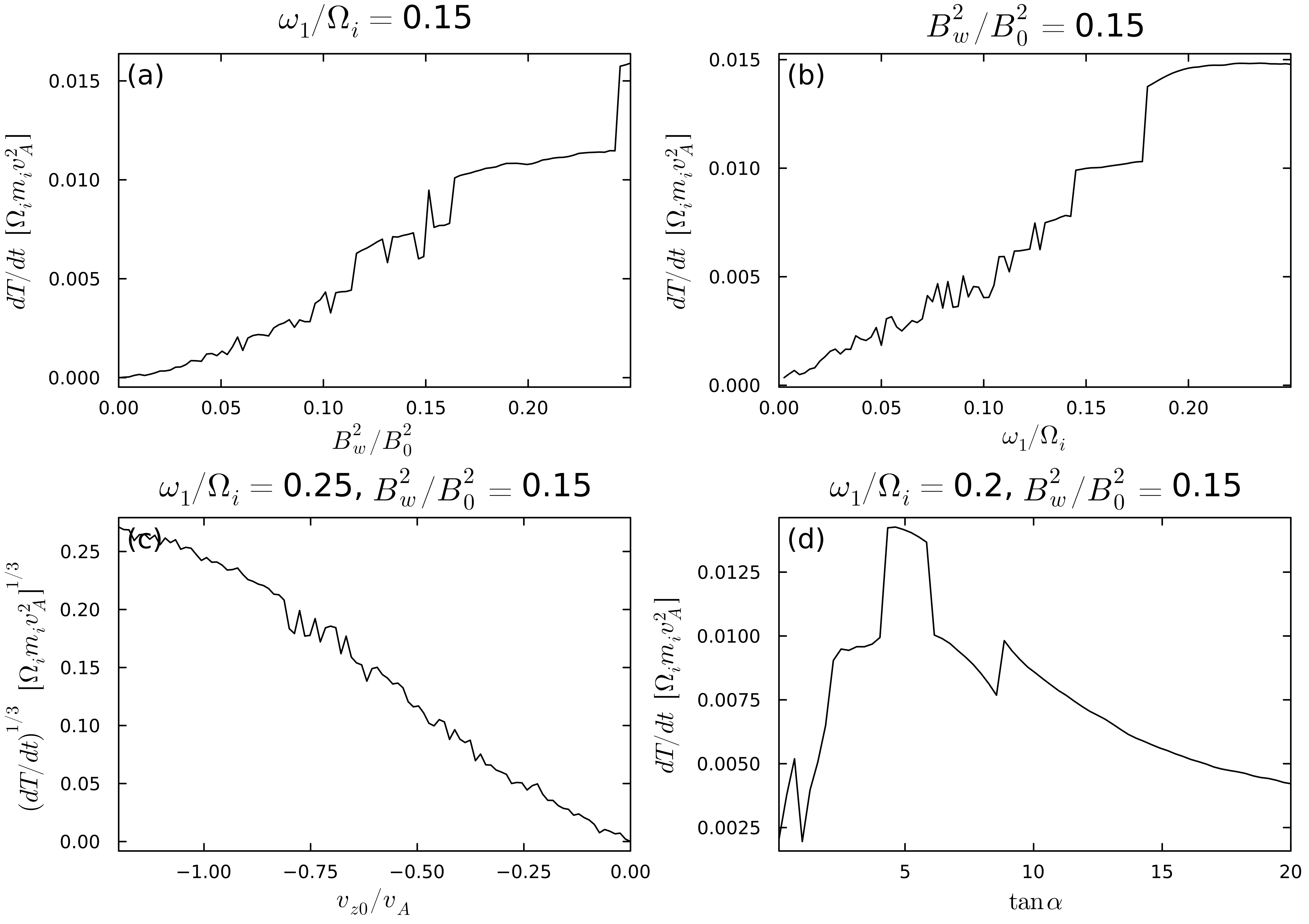}
\caption{\label{fig:heatingRate_params}The relationship between the heating rate $Q=dT/dt$ and certain quantities: (a) $B_w^2/B_0^2$  (b) $\omega_1/\Omega_i$ (c) initial velocity $v_{z0}/v_A$ (d) $\tan\alpha$. Here, $N=10$, for (a)(b)(c), $\tan\alpha=5$, and for (a)(b)(d), $v_{z0}=-v_A$.}
\end{figure}
We study how the stochastic heating rate varies with different parameters and initial velocities. The method used to calculate the heating rate $Q$ is consistent with that shown in Fig.~\ref{fig:temp_time}(a). We consider a cold plasma composed of 2000 particles, each with an initial velocity $\bm{v_0}=v_{z0}\hat{\bm{e_z}}$ and random initial phases. As shown in Fig.~\ref{fig:heatingRate_params}(a)\&(b), the heating rate increases with both $B_w^2 / B_0^2$ and $\omega_1 / \Omega_i$, demonstrating a clear linear relationship,
\begin{eqnarray}
    Q&\propto& B_w^2/B_0^2,\label{eq:Q_Bw}\\
    Q&\propto& \omega_1/\Omega_i.\label{eq:Q_omega1}
\end{eqnarray}
The heating rate and the initial velocity $v_{z0}$ exhibit a good power-law relationship, see Fig.~\ref{fig:heatingRate_params}(c),
\begin{equation}
    Q\propto \left|v_{z0}/v_A\right|^{3}.\label{eq:Q_v}
\end{equation}
This implies that as the particles' bulk velocity approaches the waves' phase speed, i.e., $\left|v_{z0}\right|\to 0$, the stochastic heating rate decreases rapidly.
Combining Eq.~\ref{eq:Q_Bw}-\ref{eq:Q_v}, the dimensionless heating rate
\begin{equation}
    \tilde{Q}=\frac{Q}{\Omega_im_iv_A^2}=H\left(\alpha\right)\tilde{v}^3\tilde{B}_w^2\tilde{\omega_1}.\label{eq:heatingRate}
\end{equation}
Where $H(\alpha)$ describes the effect of the propagation angle $\alpha$ on the heating rate, which is shown in Fig.~\ref{fig:heatingRate_params}(d). We find that the heating rate is maximum near $\tan\alpha\approx5$, corresponding to a propagation angle $\alpha$ of approximately $80^{\circ}$. 

\begin{figure}[htb]
\includegraphics[scale=0.032]{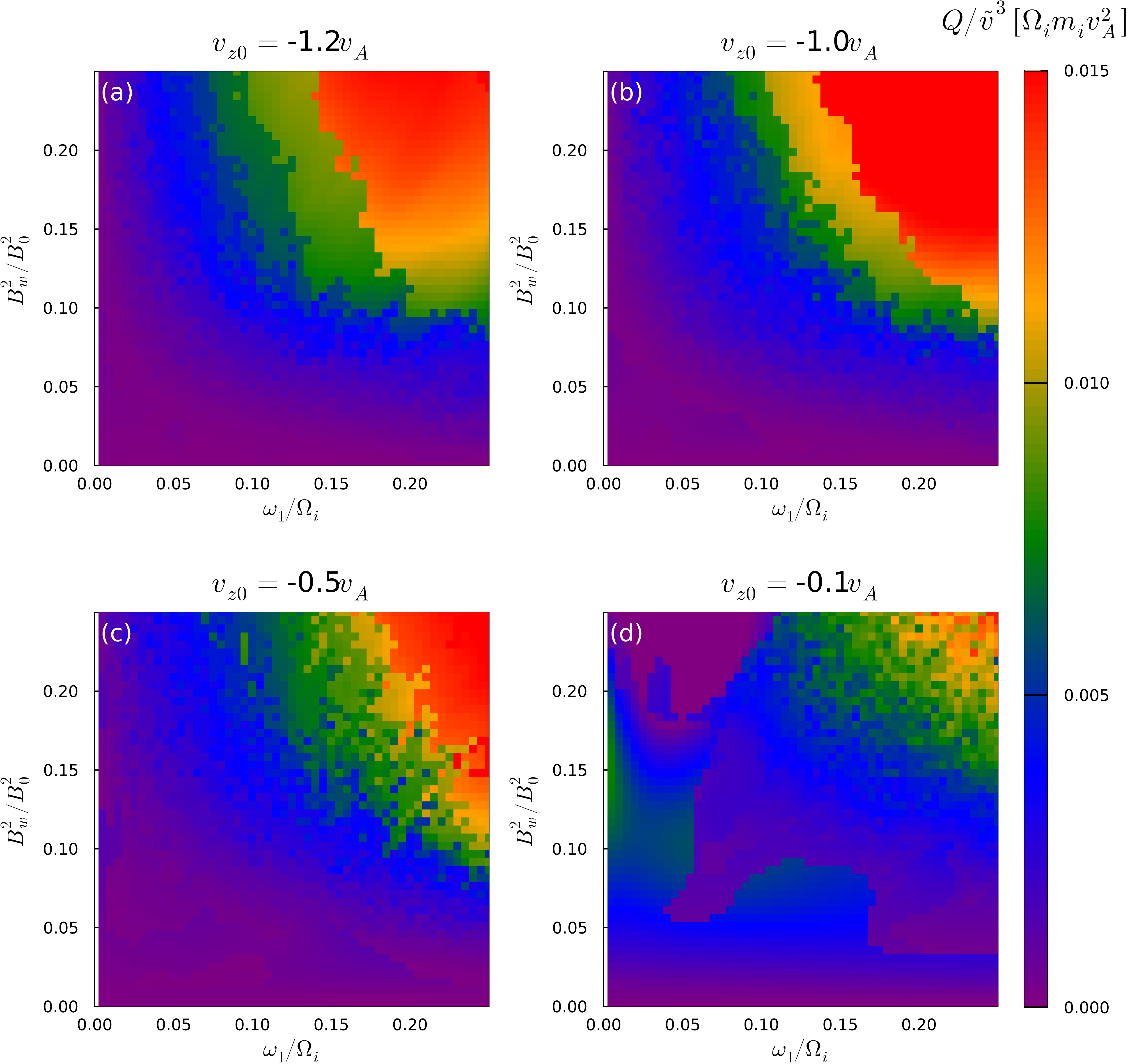}
\caption{\label{fig:heatingRate_paramspace}The stochastic heating rate $Q$ in the parameter space $(\omega_1/\Omega_i, \,B_w^2/B_0^2)$, $Q$ is normalized by $\tilde{v}^3$. $N=10$, $\tan\alpha=5$. For (a)-(d), the initial velocities are $v_{z0}/v_A=-1.2,\,-1,\,-0.5,\,-0.1$, respectively.}
\end{figure}

\begin{figure}[htb]
\includegraphics[scale=0.05]{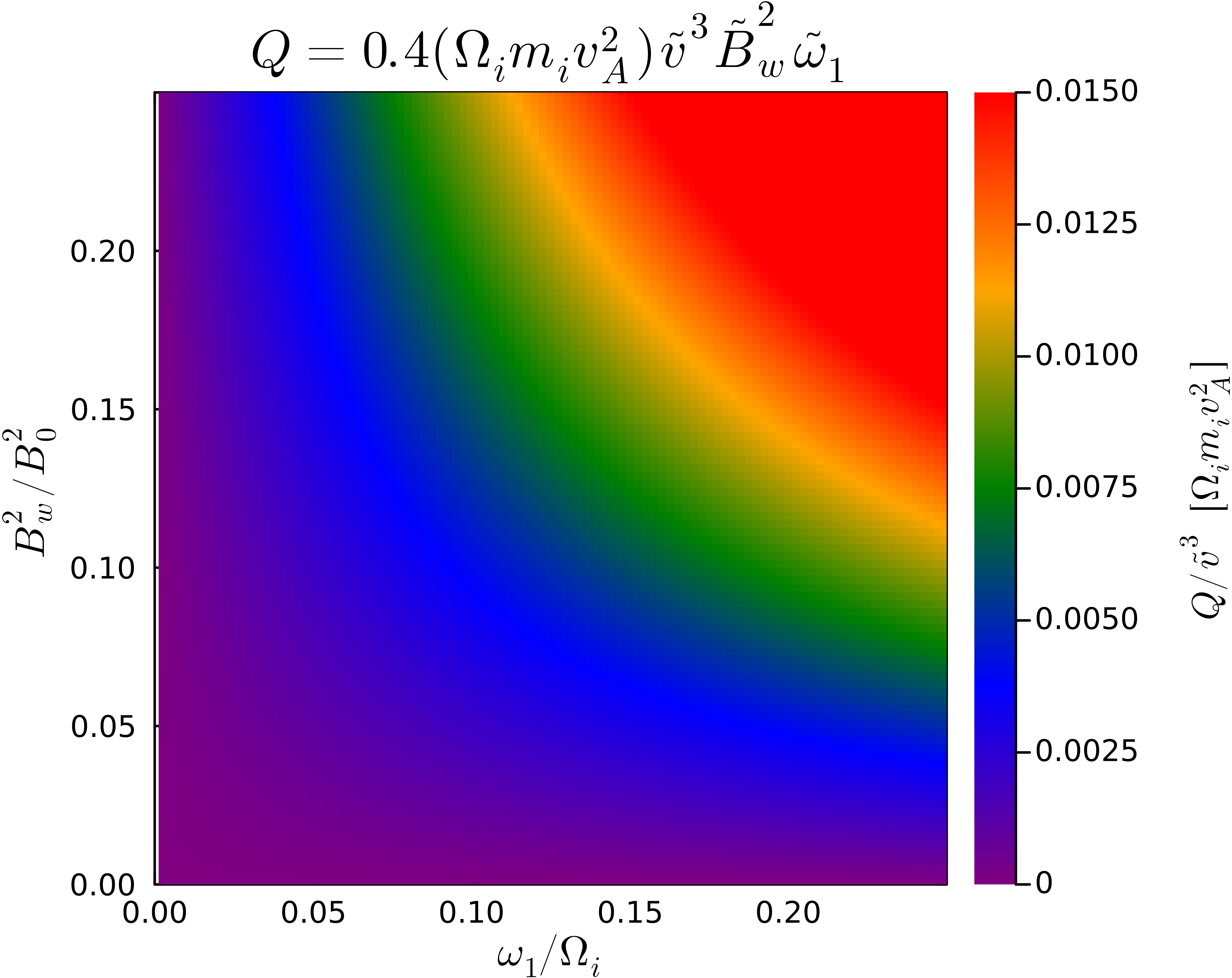}
\caption{\label{fig:heatingRate_model}The heating rate model at $\tan\alpha=5$, with colors represent the heating rate normalized by $\tilde{v}^3$.}
\end{figure}

Fig.~\ref{fig:heatingRate_paramspace} shows the heating rates in the parameter space $(\omega_1/\Omega_i, \,B_w^2/B_0^2)$ for different initial velocities $v_{z0}$, with $\tan\alpha=5$.
Based on this result, an estimate of $H(\alpha)$ at $\tan\alpha=5$ can be provided, $H\approx0.4$. Consequently, a heating rate model at $\tan\alpha=5$ can be expressed as
\begin{equation}
    Q=0.4\tilde{v}^3\tilde{B}_w^2\tilde{\omega_1}\left(\Omega_im_iv_A^2\right)\label{eq:heatingRateModel}.
\end{equation}
This heating rate model is plotted in Fig.~\ref{fig:heatingRate_model}, which is in close agreement with the particle simulation results presented in Fig.~\ref{fig:heatingRate_paramspace}.
Eq.~\ref{eq:heatingRateModel} can be regarded as a typical value for stochastic heating rates associated with quasi-perpendicular low-frequency AWs, and can be applied to calculate the stochastic heating rate in AW-turbulent plasma environments, such as the solar wind and corona.

\section{\label{sec:dis}conclusion and discussion}

We study the chaotic motion and stochastic heating of ions in the low-frequency AW spectrum using test particle simulations. The maximum Lyapunov exponent $\lambda_m$ and $CR$ are employed to quantify chaos. The chaos border in parameter space is defined by $CR=0.01$. The nature of chaotic ion motion is the breakdown of ion magnetic moment conservation caused by WFLC. Chaotic behavior can be determined by a single parameter, $P_{eff.}$. When $P_{eff.}\lesssim25$, chaotic behavior emerges. The chaos border can then be theoretically determined by $P_{eff.}=25$. The results show excellent agreement with the test particle simulation results. The minimum $P_{eff.}$ across the entire space, $P_{eff.}^m$, decreases as $N$ increases, indicating a higher likelihood of chaotic behavior. We speculate that chaotic behavior and stochastic heating are almost certain to occur in the AW continuum spectrum of the solar wind and corona.

The variation of ion temperature in the low-frequency AW spectrum is investigated. We find that as $N$ increases, the system's temperature-time curve converges, suggesting that $N=10$ would be sufficient to accurately represent the heating characteristics of the continuum spectrum. Furthermore, we observed that the heating behaviors of cold plasma ($\beta=0$) and low-$\beta$ plasma are remarkably similar.

Stochastic heating arises from the diffusion of the magnetic moment distribution caused by WFLC. It exhibits distinct anisotropic characteristics at different stages: perpendicular heating in the early stage, quasi-isotropic heating in the middle stage, and parallel heating in the final stage. A uniform solid angle distribution model is employed to qualitatively explain the anisotropic characteristics of stochastic heating.

The stochastic heating rate $Q$ is calculated, leading to the relationship between the heating rate and wave conditions and initial velocities expressed in Eq.~\ref{eq:heatingRate}. We estimate the value of $H(\alpha)$ for $\tan\alpha=5$, and this result can be used to calculate the stochastic heating rate of quasi-perpendicular low-frequency AWs in  space plasma.

To estimate the stochastic heating rate in the solar wind, we consider the inflection point between the turbulent energy-containing region and the inertial region of the solar wind AW turbulence as $\omega_1/\Omega_p \approx 10^{-4}$, with $B_w^2/B_0^2$ set to 0.1. Based on typical proton parameters at 1 au, the stochastic heating rate given by Eq.~\ref{eq:heatingRateModel} reaches $4\times 10^{-17}W/m^3$, which closely matches the heating rate results calculated from observations by the Parker Solar Probe~\cite{Wu2020}.
The heating timescale is $\tau\sim\frac{1/\Omega_i}{0.4\tilde{v}^3\tilde{B_w}^2\tilde{\omega_1}}$. For the solar wind, the characteristic heating time is approximately 100 hours. Therefore, it is highly probable that the in-situ detection data did not capture the maximum heating state of stochastic heating—where ion velocities are uniformly distributed over the spherical surface—resulting in significant perpendicular temperature anisotropy. This  anisotropy can excite instabilities that trigger ion cyclotron waves~\cite{[][{, and references therein}]Verscharen2013}, forming a joint mechanism of stochastic heating and instability. This joint mechanism could cause ion heating and turbulence cascades that transfer wave energy from large scales down to the ion gyro-scale.

The stochastic heating rate obtained from our test particle simulations differs from the AW turbulent stochastic heating rate presented in Ref.~\cite{Chandran2010}. The main difference is that their use of the linear polarization relation for velocity disturbances at the particle gyro-scale, while we determine particle velocities by solving nonlinear governing equations. A more detailed analysis of these differences will be the subject of future work.


\bibliography{pjy2025}

\end{document}